\documentclass[twocolumn,aps,english,superscriptaddress]{revtex4-1}
\usepackage[T1]{fontenc}
\usepackage[latin9]{inputenc}
\usepackage{amsmath}
\usepackage{amssymb}
\usepackage{cancel}
\usepackage{graphicx}
\usepackage{epsfig}
\usepackage{epstopdf}
\usepackage{esint}
\usepackage[colorlinks=true,citecolor=blue,urlcolor=blue]{hyperref}
\usepackage{color}
\usepackage{verbatim}
\makeatletter

\@ifundefined{textcolor}{}{%
 \definecolor{BLACK}{gray}{0}
 \definecolor{WHITE}{gray}{1}
 \definecolor{RED}{rgb}{1,0,0}
 \definecolor{GREEN}{rgb}{0,1,0}
 \definecolor{BLUE}{rgb}{0,0,1}		
 \definecolor{CYAN}{cmyk}{1,0,0,0}
 \definecolor{MAGENTA}{cmyk}{0,1,0,0}
 \definecolor{YELLOW}{cmyk}{0,0,1,0}
}

\usepackage{dcolumn}
\usepackage{bm}
\usepackage{enumerate}

\usepackage{babel}
\addto\captionsenglish{}
\renewcommand{\thefigure}{\@arabic\c@figure}

\makeatother

\begin{document}

\title{Turbulent cascade induced persistent current of cold atomic superfluids}
\author{Xiyu Chen}
\affiliation{Shaanxi Key Laboratory for Theoretical Physics Frontiers, Institute of Modern Physics, Northwest University, Xi'an 710127, China}
\affiliation{School of Physics, Northwest University, Xi'an 710127, China}
\author{Tao Yang}
\email{yangt@nwu.edu.cn}
\affiliation{Shaanxi Key Laboratory for Theoretical Physics Frontiers, Institute of Modern Physics, Northwest University, Xi'an 710127, China}
\affiliation{School of Physics, Northwest University, Xi'an 710127, China}
\affiliation{NSFC-SPTP Peng Huanwu Center for Fundamental Theory, Xian 710127, China}
\author{Wen-Li Yang}
\email{wlyang@nwu.edu.cn}
\affiliation{Shaanxi Key Laboratory for Theoretical Physics Frontiers, Institute of Modern Physics, Northwest University, Xi'an 710127, China}
\affiliation{School of Physics, Northwest University, Xi'an 710127, China}
\affiliation{NSFC-SPTP Peng Huanwu Center for Fundamental Theory, Xian 710127, China}
\author{Wu-Ming Liu}
\email{wmliu@iphy.ac.cn}
\affiliation{Beijing National Laboratory for Condensed Matter Physics, Institute of Physics, Chinese Academy of Sciences, Beijing, China}

\begin{abstract}
Dissipating of disorder quantum vortices in an annular two-dimensional Bose-Einstein condensate can form a macroscopic persistent flow of atoms. We propose a protocol to create persistent flow with high winding number based on a double concentric ring-shaped configuration. We find that a sudden geometric quench of the trap from single ring-shape into double concentric ring-shape will enhance the circulation flow in the outer ring-shaped region of the trap when the initial state of the condensate is with randomly distributed vortices of the same charge. The circulation flows that we created are with high stability and good uniformity free from topological excitations. Our study is promising for new atomtronic designing, and is also helpful for quantitatively understanding quantum tunneling and interacting quantum systems driven far from equilibrium.
\end{abstract}
\maketitle


Study of persistent flow of superfluid enables understanding of fundamental characteristics of superfluidity and may lead to applications in high-precision metrology and atomtronics\,\cite{PRL.95.143201,PRL.103.140405,nature.506.200}. Thanks to the technical development for achieving tailored trapping potential with arbitrary geometries, quantum transport experiments with quantum gases can be carried out in Bose-Einstein condensate (BEC) systems with weak interatomic interactions, which can be accurately described in the frame of the mean-field Gross-Pitaveskii (GP) equation\,\cite{RMP.73.307}. Two-dimensional (2D) annular BECs in ring-shaped traps\,\cite{PRL.95.143201,PRA.74.023617,PRL.99.260401,NJP.10.043012}, sustaining persistent currents with a quantized circulation around any closed path\,\cite{ PRL.99.260401,PRL.106.130401,PRA.86.013629,nature.506.200,PRL.111.205301} with the corresponding phase winding of the macroscopic wave function being an integer multiple of $2\pi$, attract growing interest as an atomic analog of superconducting quantum interference device constituting a basic element for atomtronic circuits\,\cite{PRA.75.023615,PRL.103.140405,NJP.19.020201,Jopt.18.093001}, and are analogous to persistent currents in superconducting rings with a quantized magnetic flux.

The direct macroscopic circulation flow of superfluid BECs in annular traps can be achieved by rotating {a perturbing potential}\,\cite{PRA.58.580,PRL.110.025302}. Although it is efficient, excitations in the condensate flow are hardly avoidable when the rotation is fast. Internal
atomic state manipulation\,\cite{PRL.99.260401,PRL.106.130401,PRL.110.025301} is another way to achieve circulation states. However, the experimental setup is difficult for magnetically trapped condensates due to the coupling of different Zeeman substates.  
Generating circulation states by phase imprinting\,\cite{PRA.86.013629,PRL.97.170406,PRA.97.043615} is also hard to be applied in systems with complex structures. Matter-wave guiding of BECs over large distances was realized experimentally in a neutral-atom accelerator ring based on magnetic time-averaged adiabatic potentials which can result in accelerating of BECs with hypersonic velocities but nonuniform density\,\cite{nature.570.205}.

Generally, adding energy into a system through transient stirring usually leads to more disorder. However, in 2D superfluids, vortex bending and tilting and Kelvin wave perturbations are suppressed. Hence, the well known inverse energy cascade\,\cite{PhysFluids.10.1417, PhysFluids.11.671, PhysFluids.12.233, RepProgPhys.43.547, PhysRep.362.1}, i.e. the emergence of order from turbulence, is induced by the inhibition of energy dissipation at small length scales\,\cite{PRL.113.165302,PRA.93.043614,Science.364.1264,Science.364.1267}. One of the most remarkable phenomena is the formation of Onsager vortex clusters\,\cite{Nuo.Cim.6.279}, which can be used as a vortex thermometry\,\cite{PRL.120.034504}. Moreover, in toroidal BECs, a large-scale persistent flow of atoms can be induced by the decay of 2D quantum turbulence (2DQT) in form of disordered vortex distributions, indicating energy transport from small to large length scales\,\cite{PRL.111.235301}.


Although the decay of turbulent cascade is studied extensively, the unpredictability inherent to turbulent systems is further confounded by physical properties such as boundaries and spatial dimensionality. It is of interest to investigate the dissipative dynamics of 2DQT with varying geometries. Moreover, by driving the systems far away from equilibrium through a sudden geometrical quench of the trapping potential, we may find out some novel nonequilibrium phenomena. In this letter, by designing a BEC system in {a double concentric ring-shaped trap}, we study quantitatively the quench dynamics of condensates with random vortex distributions and the emergence of persistent atom flows (currents) with high winding numbers in this novel geometry, which will facilitate the study of superfluidity and quantum tunneling in a large range of transport regimes.


\begin{figure}[tbp]
\includegraphics[angle=0,width=0.5\textwidth]{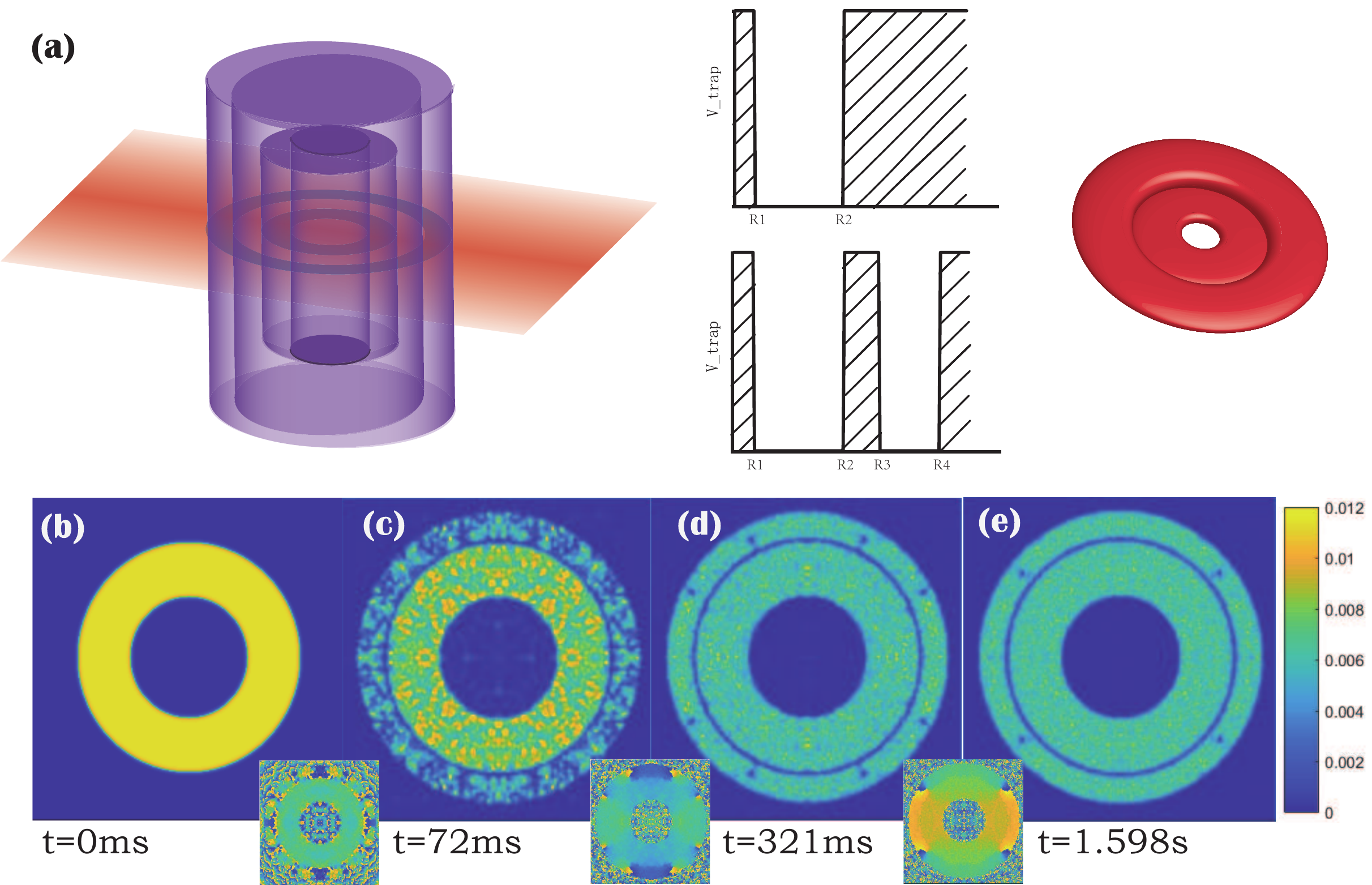}
\caption{Proposed experimental configuration (a) and some typical density distributions of the condensate with the corresponding phase diagrams during the dynamics of the system (b)-(e). The trap is deformed suddenly from the initial single ring-shaped configuration to a double concentric ring-shaped configuration at the beginning of the dynamical process. }
\label{fig1}
\end{figure}

At zero temperatures, the complex-valued mean field order parameter of a trapped 2D condensate with $N$ atoms satisfies the GP equation of the form
\begin{align}
i\hbar\frac{\partial \psi}{\partial t}=\left(-\frac{\hbar^2}{2m}\nabla^2+V_{trap}+g_{2D}N\left|\psi\right|^2\right)\psi\text{,}
\end{align}
where $g_{2D}$ is the 2D coupling constant and the order parameter $\psi$ is normalized to 1. To avoid the detrimental effects of density variation, our system is based on an experimentally accessible box potential\,\cite{PRL.110.200406,nature.539.72} where the condensates are spatially uniform away from the boundaries. Then, an 2D annular BEC is prepared in a double-concentric ring-shaped box potential of the form
\begin{align}
V_{trap}=\left\{
   \begin{array}{rcl}
 0      &    &{(R_1< r < R_{2}) \,\bigcup\, (R_3 < r < R_{4})}\\
 V_{0}  &    & elsewhere
   \end{array}\right.
\end{align}
with $r=\sqrt{x^{2}+y^2}$. {This configuration can be realized in the optical-box trap formed by two hollow tube beam and two sheet beams as shown schematically in Fig.\,\ref{fig1}(a).} $R_j\ (j=1,3)$ and $R_k\ (k=2,4)$ being the inner and outer radius of the two concentric rings, respectively. The width of the potential barrier between the two rings is then $d=R_3-R_2$. The units of time and length are $t_0=1/\omega_z$ and $a_0=\sqrt{\hbar/m\omega_z}$, respectively. If not specified, we choose $V_0=1.43\hbar\omega_z$ and $d=1.25a_0$.

In our study, the superfluid system contains a BEC of $N=10^5$ $^{87}$Rb atoms with $\omega_z=2\pi\times 700$ Hz.
The injection of angular momentum by vortex imprinting is embedded in vortex distribution rather than the macroscopic annular superflow, which is the same as in Ref.\,\cite{PRL.111.235301}. The randomly distributed vortices can be seeded in the condensate cloud by employing the phase imprinting technology or optical obstacle stirring. The initial states of the system are obtained by the imaginary time evolution of the GP equation numerically. We note that the vortices can be seeded into the system either initially or during the dynamic process, which will not affect the intrinsic physics. However, the latter will excite stronger density wave oscillations in the system.

\begin{figure}[bp]
\begin{center}
\includegraphics*[angle=0,width=0.5\textwidth]{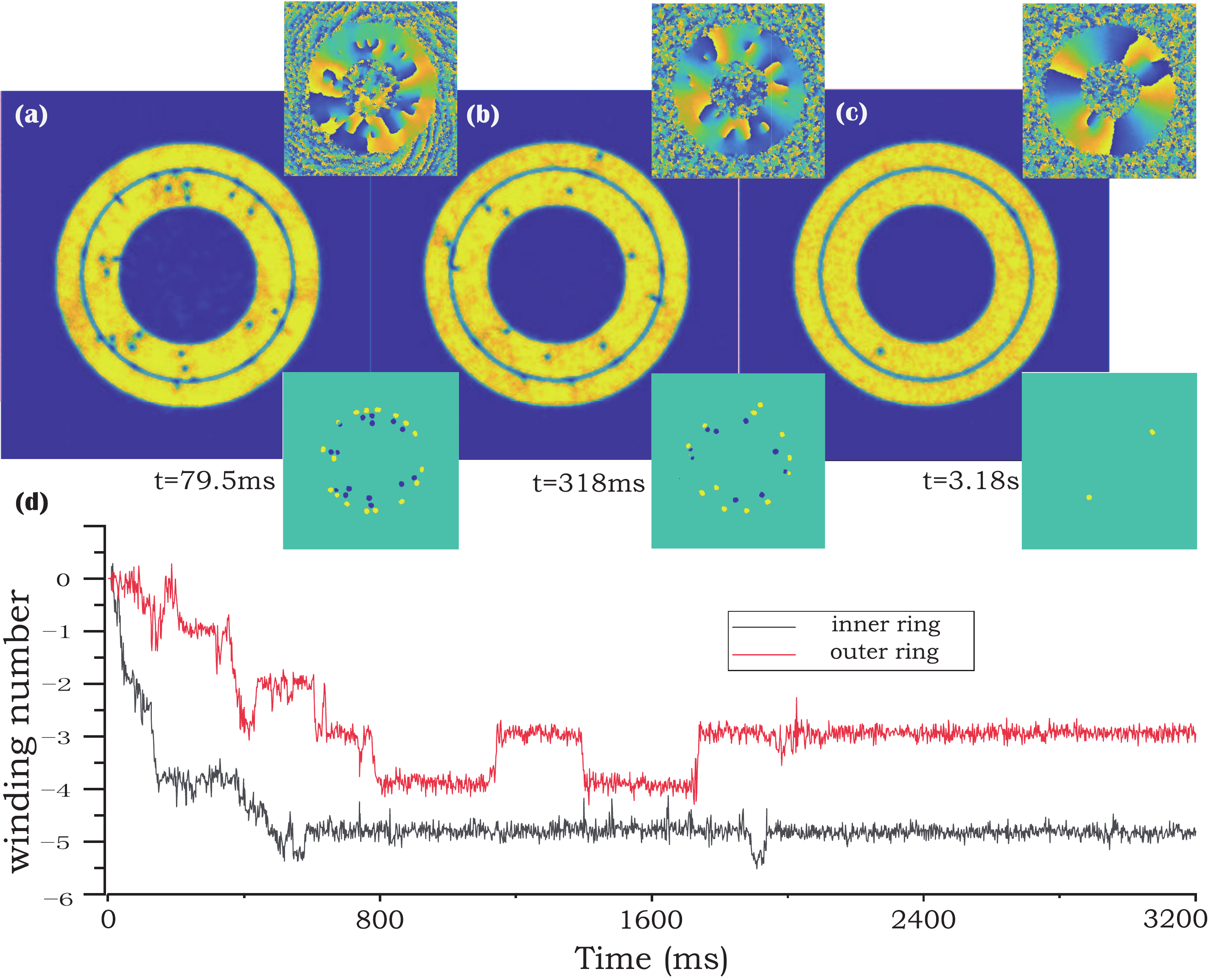}
\caption{ Typical density profiles of the evolution of the condensate initially prepared in a double concentric ring-shaped trap with corresponding phase diagrams and vortex distributions (a)-(c), and the time evolution of the winding number of the persistent current (d) with 20 initially seeded vortices.}  
\label{fig2}
\end{center}
\end{figure}

\begin{figure}[tbp]
\begin{center}
\includegraphics[angle=0,width=0.4\textwidth]{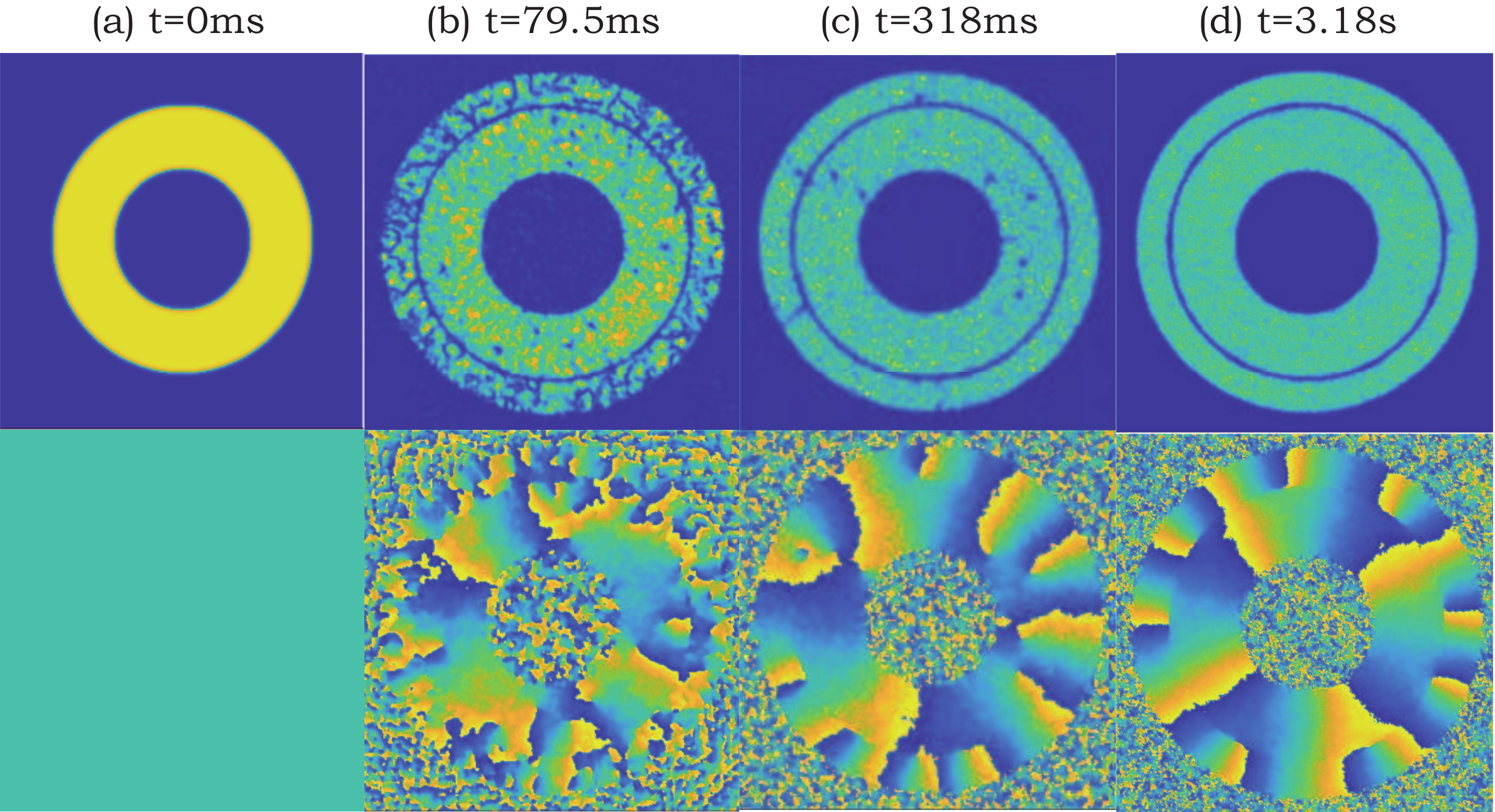}
\includegraphics[angle=0,height=0.2\textwidth,width=0.45\textwidth]{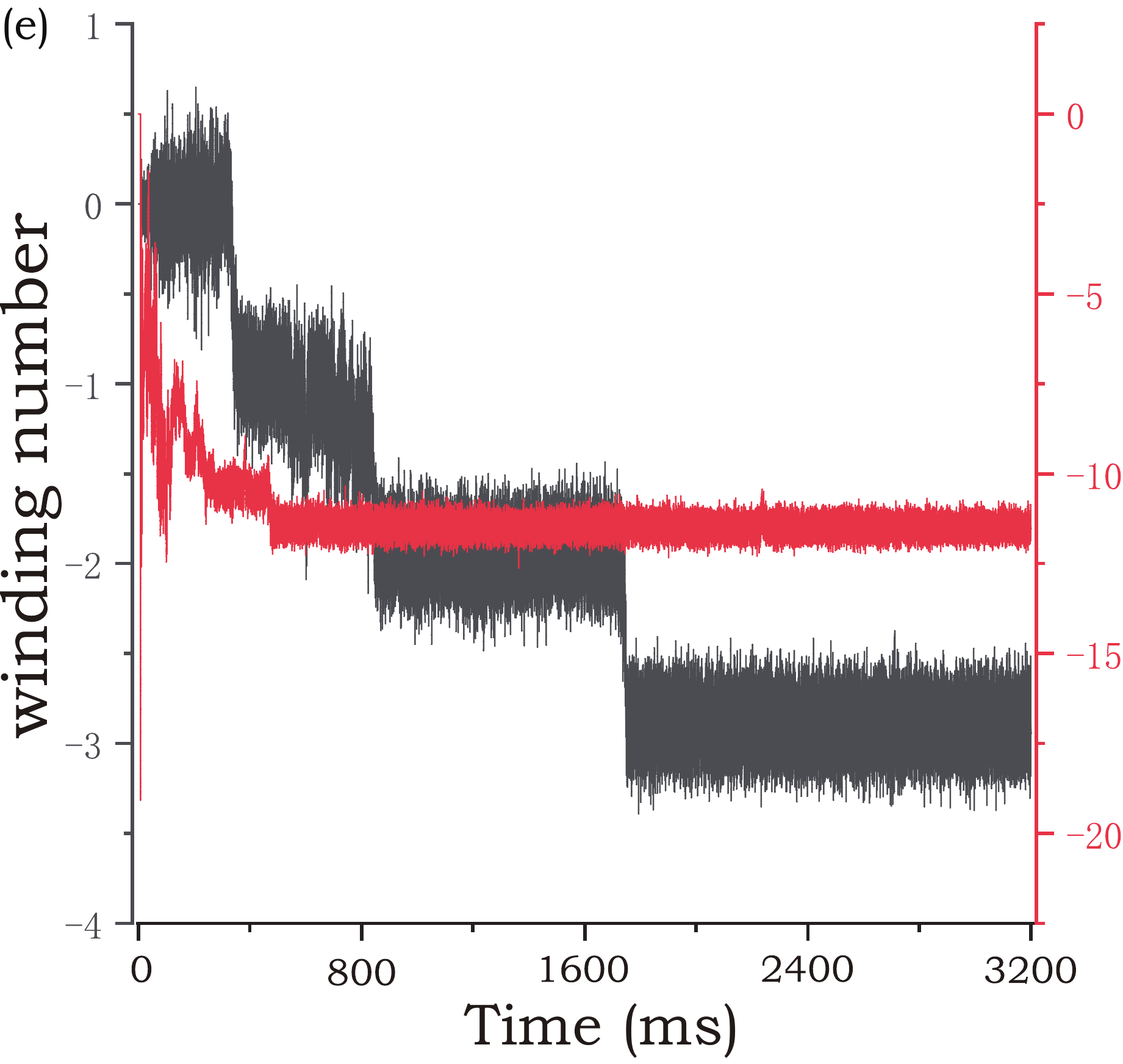}
\caption{Typical density profiles of the evolution of the condensate initially prepared in a sing ring-shaped trap and then released in a double concentric ring-shaped trap with corresponding phase diagrams (a)-(c), and the time evolution of the averaged winding number of the persistent current for 64 realizations (d) with 20 initially seeded vortices.}
\label{fig3}
\end{center}
\end{figure}

The properties of the system under geometry-quench processes away from equilibrium are less straightforward. To gain insight into the role of the quenched trapping potential, we load atoms in a single ring trap initially to obtain an annular condensate as shown in Fig.\,\ref{fig1}b, and then tune the trapping potential into a double ring geometry as schematically shown in Fig.\,\ref{fig1}a. Initially, there are no vortices and circular current in the system. After suddenly released in the double ring trap, atoms appears in the outer ring region through tunnelling across the barrier between the two rings with highly disordered density distribution as shown in Fig.\,\ref{fig1}c. With increasing homogeneity of density distribution of the condensate in the outer ring trap, vortex-antivortex pairs excited as shown in Fig.\,\ref{fig1}d without changing the total angular momentum of the system. Finally, the system reach a steady state with a few vortex pairs with symmetric distribution within the outer ring, while the density distribution of the condensate in the inner ring tends to be uniform as shown in Fig.\,\ref{fig1}e. This quadrupole structure is generally unstable in a 2D disk-shaped condensate\,\cite{SRep.6.29066}. The generation of topological excitations is similar to that during the interfering and merging of BECs in a double well potential\cite{PRA.87.023603,PRA.88.043602}. 

For condensate initially loaded in a double concentric ring trap, the ground state gives uniform density distribution in both inner and outer rings. Then, a number of vortices with the same charge, for example $s=+1$, but random positions, are excited in the inner ring region. Due to the narrow region of the ring-shaped condensate, the vortices are not far from the boundary (barrier potential between two rings). During the evolution of the system, there are ghost vortices with opposite charge emerging within the flimsy between the two parts of the condensate, whose locations are identified by the yellow points as shown in Fig.\,\ref{fig2}a-\ref{fig2}c. The number of the ghost vortices are about the same as that of the vortices. The same as previous study about 2DQT in an annular 2D quantum fluid with amount of vortices, persistent atom flow forms spontaneously in the inner ring as the number of vortices varies (decreases with respect to the initial vortex number). As shown in Fig.\,\ref{fig2}d, the current in the outer ring-shaped condensate emerges following the appearance of the current of the inner condensate. They will finally approach steady values. The variation of the circular current accompanies with the creation and annihilation of vortices in the system. The average growth or decay of the current is quantized with integer numbers indicated by the steps shown in Fig.\,\ref{fig2}, which means the transition between different quantized persistent current states. In Ref.\,\cite{PRA.80.021601}, decay of the atomic current through a barrier in a toroidal BEC was in the form of $2\pi$ sharp drops, leaving steps on the time evolution of the circulation (winding) number. 
During the time evolution, most of the ghost vortices are pinned within the barrier region and can hardly escape to the condensate cloud. But we did identify that at some stage this happened as shown in Fig.\,\ref{fig2}b-\ref{fig2}c. Moreover, the initially seeded vortices can hardly tunnel from the inner ring-shaped condensate into the outer one, and no vortices excited in the outer ring-shaped condensate comparing with the situation in Fig.\,\ref{fig1}. 

\begin{figure}[tbp]
\begin{center}
\includegraphics[angle=0,height=0.2\textwidth,width=0.45\textwidth]{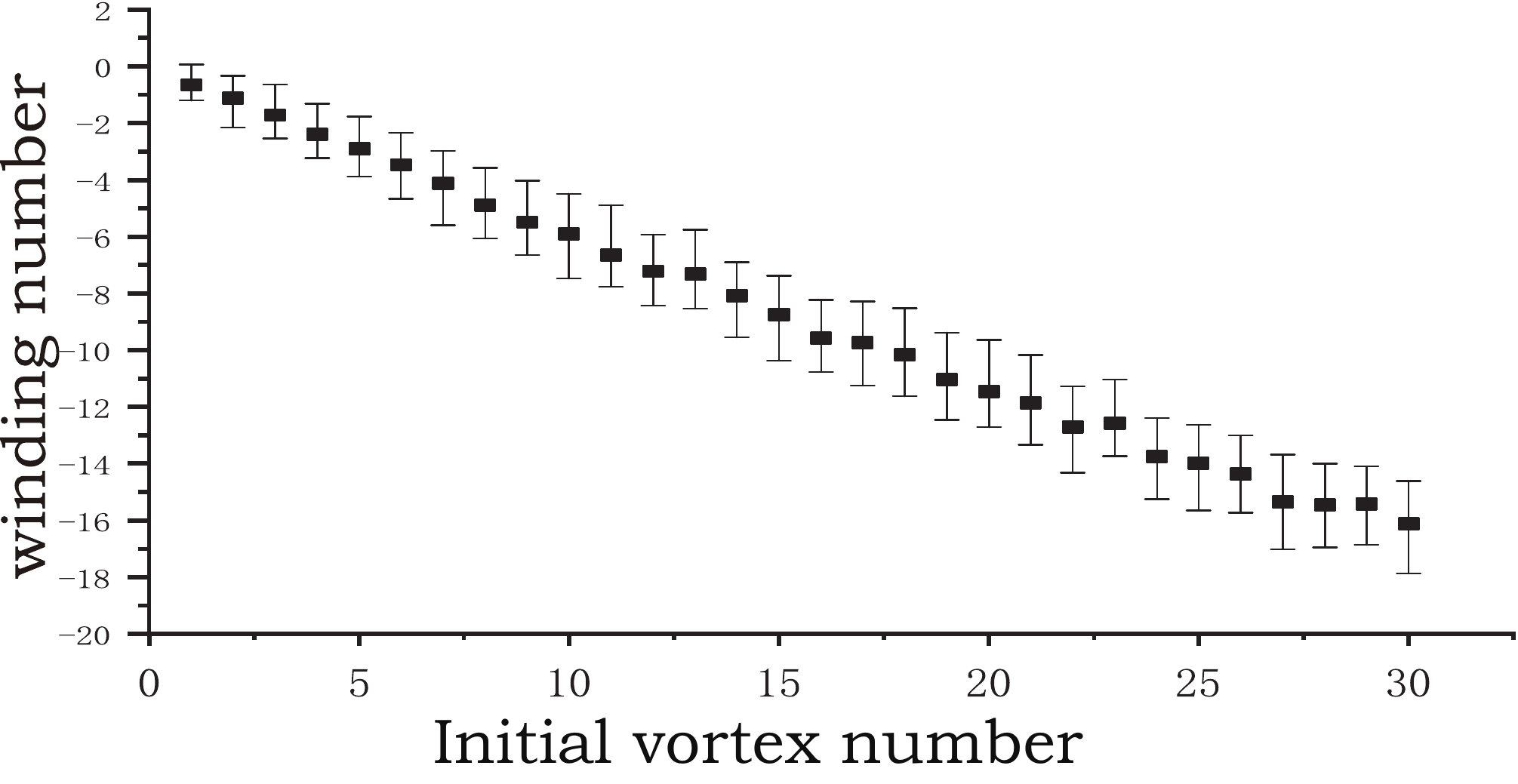}
\caption{Variation of the winding number of the circulation flow with respect to the number of initially seeded vortex. }
\label{fig4}
\end{center}
\end{figure}


Next, we set the initial state of the system to be a single ring-shaped condensate with randomly distributed vortices of the same charge, and then deform the trap into a double concentric ring-shaped geometry as what we have done for the condensate without the initial vortices in Fig.\,\ref{fig1}. The condensate originally in the inner ring tunnels through the barrier and spread in the outer ring-shaped region of the trap as shown in Fig.\,\ref{fig3}a. Formation of vortex pairs still occurs in the outer ring-shaped condensate during the tunneling process while the density distribution of the condensate is getting more and more uniform as shown in Figs.\,\ref{fig3}b-\ref{fig3}c, which is similar to the evolution of the initial state without vortices in the inner ring-shaped condensate (see Fig.\,\ref{fig1}(c)). However, different from the results shown in Fig.\,\ref{fig1}(e), all the initially seeded and dynamically excited vortices in the system decay eventually accompanied by the emergence of persistent currents in both inner and outer rings. Comparing with the result of the system without quenching of the trapping potential as shown in Fig.\,\ref{fig2}, the winding number of the current in the outer ring turns to be much larger, which means that the transfer of angular momentum from vortices to global atom flows is much more efficient in this case. 
The gray and red lines in Fig.\,\ref{fig3}g are the time evolution of the winding number of the atom flows in the inner ring and the outer ring, respectively, with a random distribution of 20 initially seeded vortices. In Ref.\,\cite{PRL.110.025302}, the circulation caused by a rotating barrier is only stable when the angular frequency of the barrier is small, while the system becomes dynamically unstable with high angular frequencies resulting from the excitations of vortices are in the condensate cloud. {The error bars show the {standard deviation} of 64 realizations for a given initial vortex distribution.} The circulation flow of the outer ring-shaped condensate reaches its steady state much faster than the inner ring-shaped condensate. In Fig.\ref{fig4}, we show that the average winding number of the outer ring-shaped condensate increases {linearly} with increasing initial number of the seeded vortices. For 30 initial vortices, the winding number of the circulation flow is -16, while that of the inner ring-shaped BEC is only -3.

\begin{figure}[tbp]
\begin{center}
\includegraphics[angle=0,height=0.2\textwidth,width=0.45\textwidth]{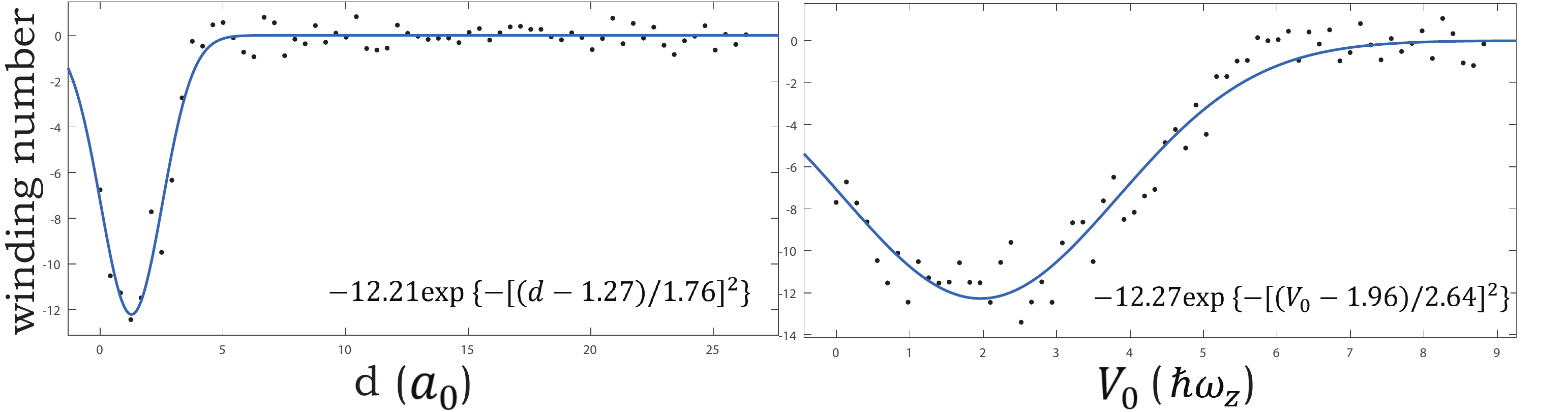}
\caption{Variation of the winding number of the circulation flow with respect to the number of initially seeded vortex. }
\label{fig5}
\end{center}
\end{figure}

We also investigate the effects of the thickness and hight of the middle flimsy barrier on the winding number of the circulation flow as shown in Fig.\ref{fig5}. The number of the initial vortices is kept to be 20. {For $d$ larger than $5a_0$} or {$V_0$ larger than 6$\hbar\omega_z$}, the tunneling is strongly suppressed to prevent the emergence of the circulation flow in the outer ring-shaped trap. The winding number of the outer ring-shaped condensate does not vary monotonically with respect to $d$ and $V_0$. There exists a specific value of these parameters to obtain maximum persistent current of the circulation flow. The variations of the winding number with respect to both parameters satisfy Gaussian distributions.

{In conclusion, we propose a novel setup and a quench protocol which can be used to generate persistent currents of circular condensate flows with high winding numbers. For a double ring-shaped condensate, if we quench vortices with the same charge and random distributions, the circular condensate flows can emerge in both inner and outer ring-shaped condensate due to 2D turbulent cascade and quantum tunneling. The differences of the winding numbers of the two flows are  not significant. If we quench the condensate with randomly distributed like vortices by deforming the trap from a single ring-shaped geometry into a concentric double ring-shaped one, the winding number of the outer ring shaped condensate will be much larger than that of the inner ring-shaped condensate flow. However, if the initial state of the system is without any vortices, the geometric quench itself cannot generate persistent atom flows in the system. The persistent currents generated in our system are with nearly uniform density distributions and free from vortex excitations. This method is much more efficient to create circulation flows of atoms than the technique of rotating barrier and may be used to design complex devices based on ultracold atoms. }

\section*{Acknowledgments}

This work is supported by the National Natural Science Foundation of China under grants Nos. 11775178, 11947301 and 61835013, National Key R\&D Program of China under grants No. 2016YFA0301500, the Strategic Priority Research Program of the Chinese Academy of Sciences under grants Nos. XDB01020300 and XDB21030300, the Major Basic Research Program of Natural Science of Shaanxi Province under grants Nos. 2017KCT-12 and 2017ZDJC-32. This research is also supported by The Double First-class University Construction Project of Northwest University.


%

\end{document}